\documentclass[amsmath,amssymb,aps,floats,prx,twocolumn,showpacs,superscriptaddress]{revtex4-1}

\usepackage{graphics}
\usepackage{graphicx}
\usepackage{amssymb}
\usepackage{epstopdf}
\usepackage{color}

\DeclareGraphicsExtensions{.pdf, .png, .jpg, .eps}
\usepackage{floatrow}

\newcommand{\iee}{{\it{i.e.}}}

\newcommand{\bfig}{\begin{figure}}
\newcommand{\efig}{\end{figure}}

\allowdisplaybreaks

\usepackage{float}

\newcommand{\bea}{\begin{eqnarray}}
\newcommand{\ena}{\end{eqnarray}}
\newcommand{\bee}{\begin{equation}}
\newcommand{\ene}{\end{equation}}

\newcommand{\muh} {\mu_0H}
\newcommand{\mub} {\mu_{\rm B}}

\newcommand{\chiab} {\chi_{ab}}
\newcommand{\chic} {\chi_{c}}

\newcommand{\rmeff}{\rm{eff}}
\newcommand{\Hpab} { H\parallel ab}
\newcommand{\hsat} { \mu_0H_{\rm sat}}
\newcommand{\Hpc} { H\parallel c}
\newcommand{\HKc} { H_{\rm Kc}}
\newcommand{\taag} {TbAuAl$_4$Ge$_2$}
\newcommand{\gaag} {GdAuAl$_4$Ge$_2$}

\newcommand{\mab} {M_{ab}}
\newcommand{\mc} {M_c}
\newcommand{\msat} {M_{\rm sat}}
\newcommand{\tone} {T_{\rm N1}}
\newcommand{\ttwo} {T_{\rm N2}}
\newcommand{\tthree}{T_{\rm N3}}
\newcommand{\cmag} {C_{\rm mag}}
\newcommand{\smag} {S_{\rm mag}}
\newcommand{\Dcmag} {\Delta C_{\rm mag}}
\newcommand{\dmdh} {\frac{dM_{ab}}{dH}}

\begin{document}

\title { Field-Induced Magnetic States in the Metallic Rare-Earth Layered Triangular Antiferromagnet TbAuAl$_4$Ge$_2$}

\author{Ian A. Leahy}
\affiliation{Department of Physics, University of Colorado, Boulder, Colorado}

\author{Keke Feng}
\affiliation{Department of Physics, Florida State University, Tallahassee, Florida}

\author{Roei Dery}
\affiliation{Department of Physics, Cornell University, Ithaca, New York }

\author{Ryan Baumbach}
\affiliation{Department of Physics, Florida State University, Tallahassee, Florida}
\affiliation{ National High Magnetic Field Laboratory, Tallahassee, Florida }

\author{Minhyea Lee}
\affiliation{Department of Physics, University of Colorado, Boulder, Colorado}

\date{\today}

\begin{abstract}
Magnetic frustration in metallic  rare earth lanthanides ($Ln$) with $4f$-electrons   is crucial for producing interesting magnetic phases with high magnetic anisotropy  where intertwined charge and spin degrees of freedom lead to novel phenomena.
Here we report on the magnetic, thermodynamic, and electrical transport properties of \taag. 
Tb ions form 2-dimensional triangular lattice layers which stack along  the crystalline $c$-axis.  
The magnetic phase diagram reveals multiple nearly degenerate ordered states upon applying field along the  magnetically easy $ab$-plane before saturation.
The magnetoresistance  in this configuration  exhibits intricate  field dependence that closely follows  that of the magnetization  while the specific heat reveals a region of highly enhanced entropy,  suggesting the possibility of a non-trivial spin textured phase.
For fields applied along the $c$-axis (hard axis), we find linear magnetoresistance over a wide range of fields.  We compare the magnetic properties and magnetoresistance with an isostructral  \gaag~single crystals. These results identify \taag~ as an environment for complex quantum spin states and pave the way for further investigations of the broader $Ln$AuAl$_4$Ge$_2$ family of materials.

\end{abstract}

\maketitle

\section {Introduction}

Geometrically frustrated lattices (e.g., triangular, Kagome, Shastry-Sutherland, and pyrochlore)
 continue to attract interest because they host a variety of strongly correlated quantum phases~\cite{Balents2010, qimao2010}. The simplest examples are electronic insulators, 
 where lattice-based frustration is not disrupted by coupling between charge and spin degrees of freedom. 
 Beyond this, a variety of frustration-induced effects have been discussed in metallic magnets, 
 including flat bands \cite{Kang2020_natcomm, Kang2020_natmat}, quantum Hall effects 
 in Kagome systems \cite{Tang2011}, and exotic superconductivity \cite{Yang2008}. 
 Amongst such materials, $f$-electron metals are known for hosting  novel phenomena where correlation effects  manifest in diverse ways ~\cite{Lacroix2010,Lohneysen2007_RMP, Fushiya2014}.

Recent investigations of the centrosymmetric triangular lattice magnets Gd$_2$PdSi$_3$ \cite{Kurumajii2020} and Gd$_3$Ru$_4$Al$_{12}$ \cite{Hirschberger2019} have shown topological Hall effects associated with Skyrmion-lattice-like  spin textures in the absence of Dzyaloshinskii-Moriya interactions \cite{Muhlbauer2009}. 
 Theoretical studies have ascribed these behaviors to the competition amongst 
the Ruderman-Kittel- Kasuya-Yosida (RKKY) interaction, the Kondo effect, and the strength of the geometric frustration \cite{Doniach1977, Coleridge1987, Lucas2017}.  
More broadly, the interactions between  rare earth magnetic ions and conduction electrons provide a unique route for realizing strong electronic correlation effects  \cite{Tsuda2018, Lohneysen2007_RMP, qimao2010}. Understanding these effects is key to uncovering many other novel phenomena. 

\begin{figure}[t]
\includegraphics[width=\linewidth]{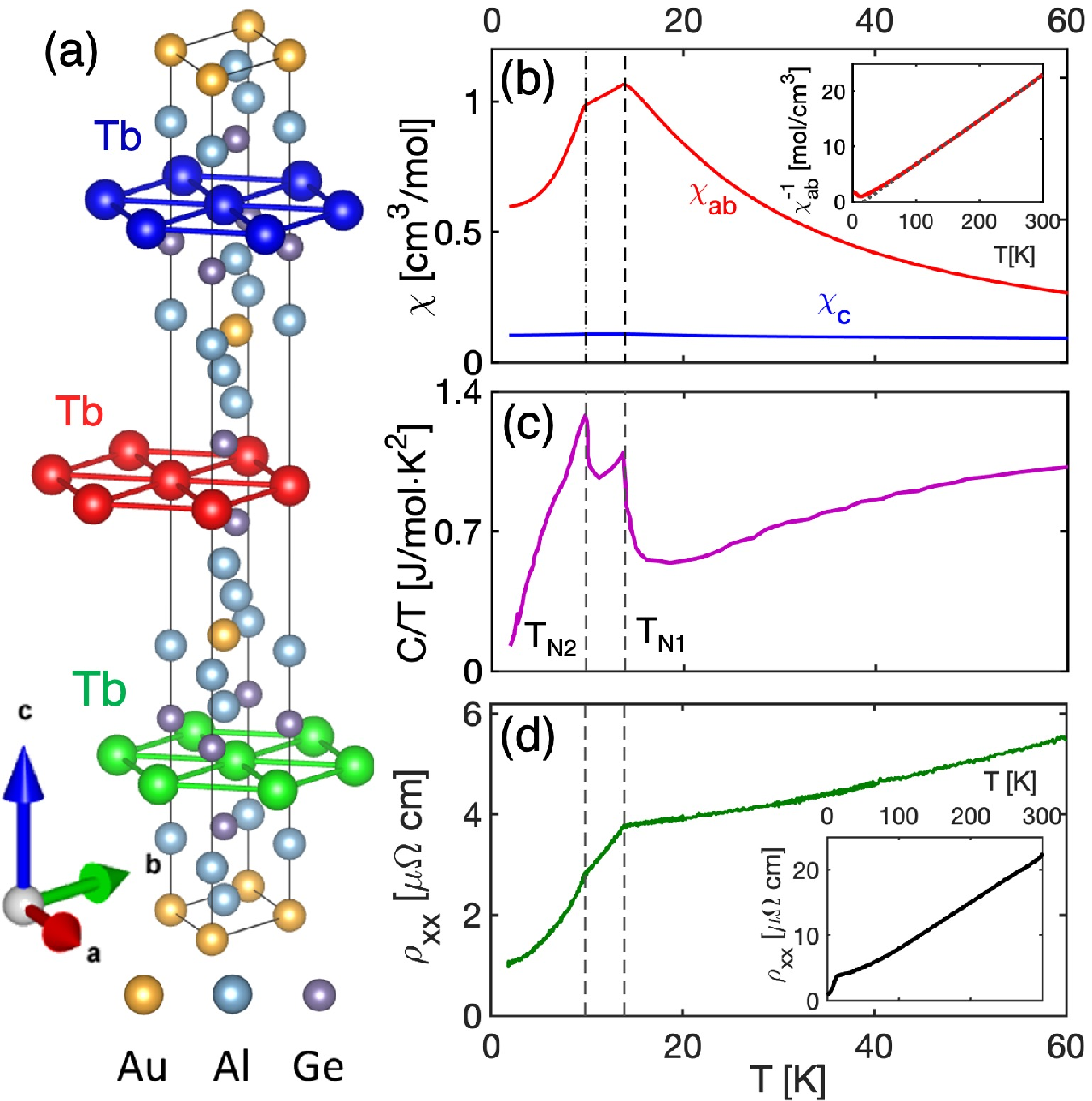}
\caption{ (a) Crystal structure of \taag. Tb ions form a 2D triangular lattice with layers stacked along the c-axis separated by  approximately 1.1nm.  
(b) Temperature dependent magnetic susceptibility measured  at $\muh = 0.5$ T for $\Hpab$ ($\chiab$) and $\Hpc$ ($\chi_c$).
The inset displays the Curie-Weiss fit of $\chiab^{-1}$ above 150 K, yielding the effective magnetic moments $9.9~\mu_B$/Tb. 
Temperature dependence of  specific heat (c)   and resistivity (d) at zero field  are shown respectively.  Two magnetic transitions at $\tone = 13.9$ K and $\ttwo= 9.8$ K are indicated  with broken  vertical lines in (b-d).}
\label{fig1}
\end{figure}

Here we present  the magnetic and electrical properties of  metallic \taag{} as a function of magnetic field and temperature. 
\taag~ crystallizes in a rhombohedral structure with staggered triangular nets of Tb$^{3+}$ ions as shown in Fig. \ref{fig1}(a). Single ions of Tb$^{3+}$ have total angular moment $J=6$ ($L=3$ and $S=3$). We study the magnetoresistance (MR) and find  starkly contrasting behavior for fields applied parallel to the Tb-layers ($\Hpab$) or perpendicular to the Tb-layers ($\Hpc$). Bulk magnetization, heat capacity, and electrical transport measurements reveal  highly anisotropic and complex magnetic ordering with several distinct regions in the temperature-magnetic field phase space. In particular, we observe nearly degenerate ordered ground states for fields applied along the magnetically easy $ab$-plane. While this behavior is not unusual for $f$-electron metals, an analysis of the specific heat uncovers a region of enhanced magnetic entropy in some of these phases, which implies the presence of nontrivial spin textures similar to those seen for the prototypical Skyrmion lattice systems MnSi~\cite{Bauer2013} and MnGe\cite{Fujishiro2018}. For fields applied along the magnetically hard $c$-axis, low-temperature electrical transport measurements additionally uncover linear magnetoresistance over a wide range of fields. Such behavior is typically associated with topological band structures~\cite{LYe2018, Leahy2018} or partial gapping of the Fermi surfaces by density waves~\cite{Feng2019,Kolincio2020}. Finally, we compare the MR results and magnetic properties with those in the isostructural \gaag{}, which shows similar behavior for fields applied along and perpendicular to the Gd-layers. Our results invite further investigations of \taag~ to expose the magnetic order parameters and their impact on electronic quantities. We anticipate that the broader $Ln$AuAl$_4$Ge$_2$ family of materials will provide a new reservoir for related quantum spin states and novel phenomena.

\section {Experimental Details}

Single crystals of \taag~and \gaag~ were grown using a molten aluminum flux method. Details on the crystal growth conditions are  found  elsewhere \cite{SZhang2017, KekeFeng2022}. 
The  space group of the crystals was confirmed  $R\bar3m$, using powder x-ray diffraction measurements with a Cu K$\alpha$ source.
Detailed structure analysis and comparison with other lanthanide element substitutions are reported in Ref. \cite{KekeFeng2022}. 
Specimens were aligned by single-crystal X-ray diffraction measurements with graphite monochromated Mo K$\alpha$ radiation. 
Temperature and magnetic field-dependent magnetization measurements were performed for $T = 1.8- 300$ K 
and $\muh = 0 - 7$ T using a Quantum Design Magnetic Properties Measurement System. 
Specific heat  measurements were performed for $T = 1.8 - 70$ K and $\muh= 0 - 5$ T using a Quantum Design Physical Properties Measurement System using a conventional thermal relaxation technique.
Electrical transport measurements were performed using a standard  low-frequency ac technique  with a four-probe  configuration  in a commercial cryostat and superconducting magnet. 
The dimensions of the transport samples used here are typically  $\sim$ 0.7$\times$0.3$\times$0.2 mm$^3$ for 
\taag~and $\sim0.5\times$0.2$\times$0.1 mm$^3$  for \gaag. 
Field symmetrization is performed on time-reversed field sweeps (up and down) to calculate the longitudinal resistivity ($\rho$)  and eliminate electrical pickup from contact misalignment.  The magnetic field was applied along either the $ab$-plane ($\Hpab$) or the $c$-axis ($\Hpc$) as in the magnetization measurements, with the current applied in the $\langle100\rangle$ Tb layer of the $ab$-plane.  For both cases, the current  lies within the $ab$-plane and perpendicular to the field direction.

\section {Results and Discussion}

\begin{figure}[t]
\includegraphics[width=\linewidth]{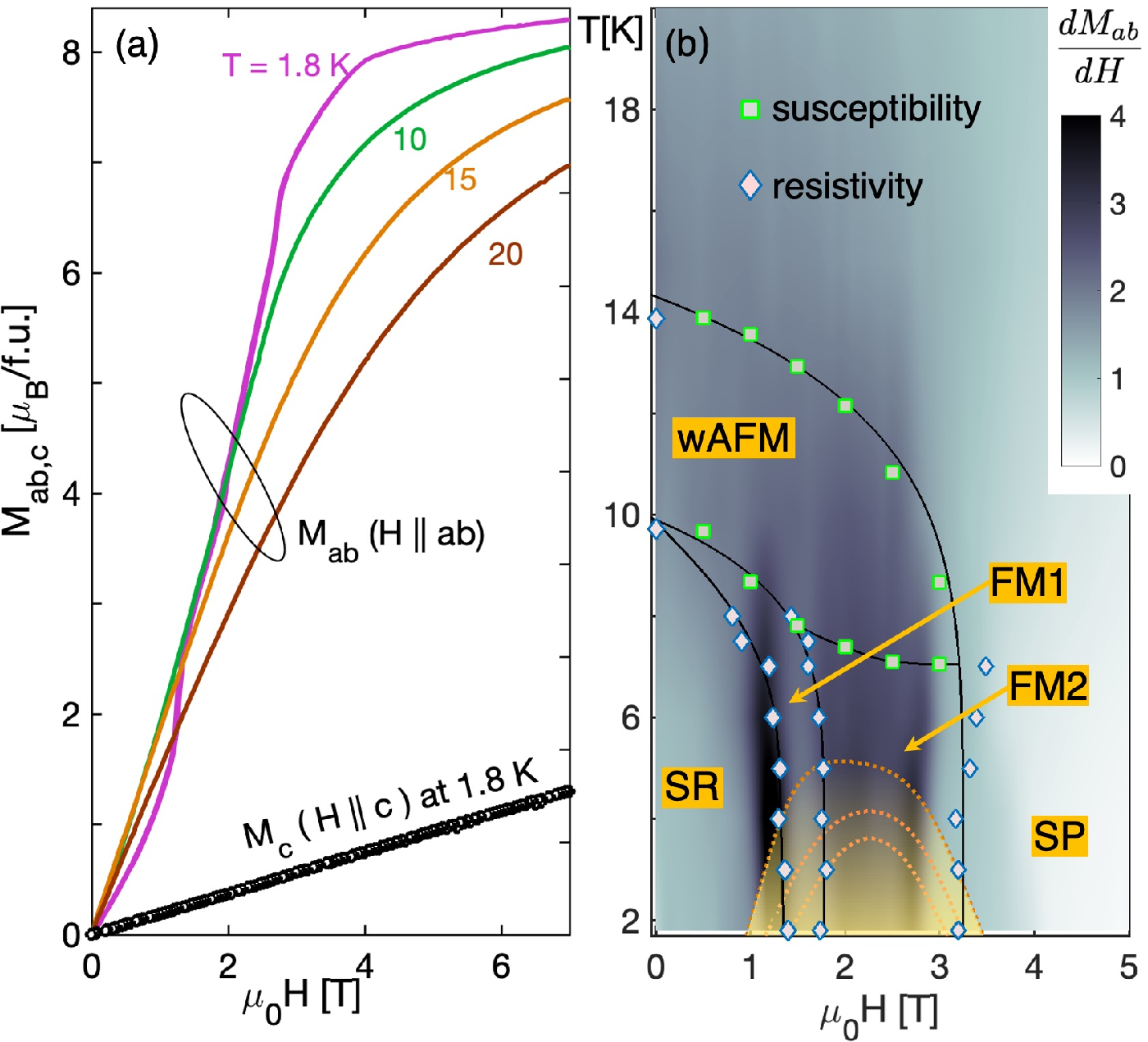}
\caption{ (a)  $\mab (H)$ (solid lines) of \taag~with field applied along the $ab$-plane. 
 At $T=1.8$ K, $\mab$ saturates at $\msat = 8.3 \,\mub$/Tb,  making the $ab$-plane 
  the magnetic easy plane.  
  For comparison, $\mc (H)$ with  $\Hpc$ at 1.8 K is displayed as black circles.
 $\mc$ increases linearly to $1.7\mub$/Tb at 7 T with no obvious features present. 
(b)  Phase diagram of \taag, where the colorbar represents the magnitude of $d\mab/dH$, 
which  follows the peak and kink positions of susceptibility and magnetoresistance.  The yellow-shaded area is adopted from  the specific heat  data, where the magnetic contribution to the specific heat and hence magnetic entropy, displays significant enhancement (See Sec.\ref{sec:HC}). Black lines are used as a guide.}
\label{fig:magpha}
\end{figure}

\subsection {In-plane Magnetization and Magnetic Phase Diagram  of \taag}

Fig. \ref{fig1}(a) shows the unit cell of \taag, where the Tb ions form staggered layers of two-dimensional triangular nets that are stacked along the $c$-axis
The Tb layers are spaced approximately 1.1 nm apart from one another and  the  in-plane nearest Tb neighbor distance is found at 0.42 nm \cite{KekeFeng2022}. 
The triangular arrangement of Tb ions is the basis of the geometric conditions for magnetic frustration, where the long-range Ruderman-Kittel-Kasuya-Yosida (RKKY) interaction and crystal electric field splitting of the Hund's rule multiplet  of Tb$^{3+}$
 are also anticipated to play an important role, adding complexity  to the landscape of  relevant energy scales. 
 As shown in Fig. \ref{fig1}(b), $\chi(T)$ shows strong anisotropy for magnetic fields applied in the $ab$-plane ($\chiab$) and along the $c$-axis ($\chic$), where $\chiab\gg \chic$. 
A Curie-Weiss fit to $\chi_{\rm{ab}}$ for $T$ $>$ 200 K yields a Curie-Weiss temperature  $\Theta_{\rm CW} = 18$ K and an effective magnetic moment $\mu_{\rmeff}=9.9~\mub$/Tb, in  a good agreement with the calculated moment value of single ion Tb$^{3+}$, 9.7 $\mub$. 
 Two  antiferromagnetic-like  orderings are identified by the  pronounced cusps in $\chiab(T)$  at   $\tone = 13.9$ K and $\ttwo = 9.8$ K, as marked with vertical dashed lines. 
 Both  transitions at $\tone$ and $\ttwo$ are aligned with peaks in the specific heat as a function of temperature $C (T)$ measured at zero field [panel (c)]. In the zero field resistivity $\rho(T)$ [panel (d)], the magnetic ordering appears as cusps corresponding to a reduction in scattering of conduction electrons upon ordering.

To further distinguish between different ordered states,  we plot the isothermal field-dependent magnetization for $\Hpab$ ($\mab$)  [Fig. \ref{fig:magpha}(a)]. The saturated magnetization at $T=1.8$ K reaches  $\msat \approx 8.3 \,\mu_B$/Tb   at $\muh  \ge 4.2 $T, where $\mab$ slowly  approaches to the value of  free  Tb$^{3+}$ ion moment,  9.7 $\mu_B$/Tb, corresponding to  the total angular  momentum $J=6$.   $\mc$ in $\Hpc$  configuration  increases linearly and only reaches 1.25$\mu_B$/Tb at 7 T and the slope of $\frac{d\mc}{dH}$ shows little temperature dependence.  

For $T<\ttwo$, $\mab$  displays distinct features that  indicate  a series  of field-induced transitions. Combined with $\chiab$ we generate a phase diagram with four distinct regions:
(i) the spin rotation region (SR), where $\mab$ monotonically increases with increasing $H$;
(ii) the first ferromagnetic region (FM1), which is bounded by a step-like increase  of $\mab$ at  $\muh =1.1$ T and a subtle inflection point at around  1.8 T;
(iii) the second ferromagnetic region (FM2), where the field derivative of $\mab$ remains larger before abruptly decreasing at the saturation field $\hsat= 3.4$ T;  
and (iv) and  the spin polarized state  (SP) characterized  with mostly saturated $\mab$ with a small $\dmdh$. 
The color scale of Fig. \ref{fig:magpha}(b) shows the magnitude of $\dmdh$ overlaid with the characteristic features marked from $\chiab(T)$ at different fields (squares) and $\rho(H)$ at  different temperatures  (diamonds). It naturally follows the boundaries of the states described above. When $ \tone<T<\ttwo$,  two  different magnetic states are distinguished only by the locations of two cusps in $\chiab(T)$ at different fields \cite{KekeFeng2022}. 
The yellow-shaded area  extending across the low-temperature region of FM1 and FM2 is derived from the region of elevated magnetic heat capacity (and thus magnetic entropy). We will discuss this in the next section.

\subsection {Specific Heat of  \taag }
\label{sec:HC}

\begin{figure}[t]
\includegraphics[width=\linewidth]{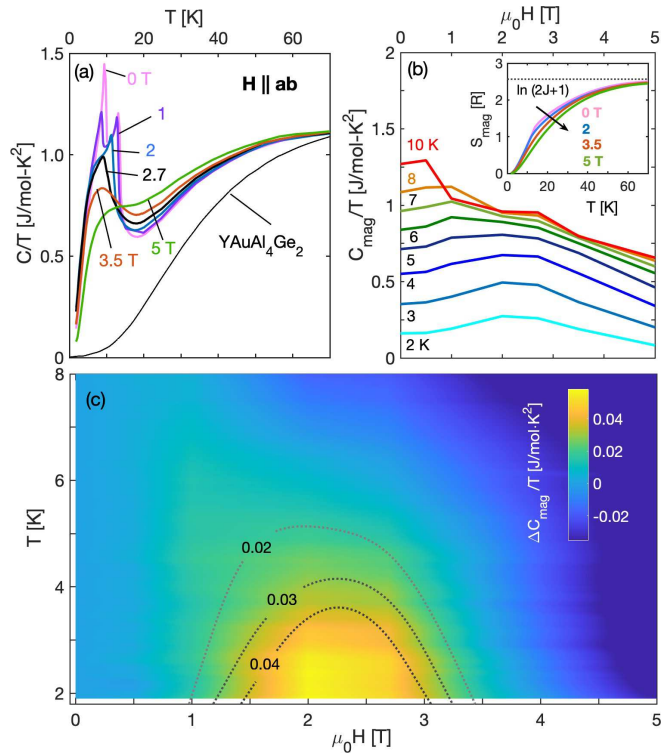}
\caption{ (a) $T$ dependence of $C/T$ at different fixed fields applied along the $ab$-plane. For $0$T, two transitions marked by peaks are visible at $\tone$ and $\ttwo$ which become broadened with field.  
(b) The magnetic field dependence of $\cmag/T$ plotted at $T=2$-10 K in 1 K increments. A shallow maximum in  $\cmag(H)$ fades away in $T\ge 6$ K.  
Inset: Magnetic entropy, $\smag(T)$, at different fixed fields calculated using $\partial S_\textrm{mag}/\partial T= C_{\rm mag}/T$. For all fields, $\smag$ converges towards $R\ln (2J+1)$, where $J =6$ for the total angular momentum of Tb$^{3+}$.
 (c) $\Dcmag(H)/T$ defined as defined as $[\cmag(H) - \cmag(H=0)]/T$ is plotted in color in $H-T$ space, which visualizes  the area of  enhanced $\Dcmag$ centered on  the low $T$ region of FM1 and FM2 states. }
\label{fig:hc}
\end{figure}

Fig.\ref{fig:hc}(a) shows the plot  of specific heat $C (T)$ at different fields applied along the $ab$-plane. At zero field,  two  transitions at $\tone$ and $\ttwo$ are marked by two peaks.  
Both peaks move to lower temperatures $T$ with increasing field but the peak at $\ttwo$ is completely suppressed by $\muh = 2$ T. 
Enhancement of $C/T$ in higher temperature  ($T>\tone$) persists for all fields, which makes it difficult to observe the phase boundaries at FM1 and FM2. 
$C(T)$ of  the non-magnetic analog of YAuAl$_4$Ge$_2$ is shown as well, which was used to obtain the magnetic contribution $\cmag = C-C_{\rm YAAG}$.  
In Fig. \ref{fig:hc}(b) we plot $\cmag (H)/T$ at several fixed temperatures: a substantial enhancement of $\cmag (H)/T$ emerges  in the field range corresponding to FM1 and FM2 for $T < 6$ K.

In the inset of panel (b) we plot $\smag (T)$, calculated from  $\smag (T) =\int \cmag/T~ dT$. $\smag (T)$ reaches the value 20.7 J/mol-K at 70 K, asymptotically approaching  the expected value for the  total angular momentum   $J=6 $  for Tb ion multiplet $R\log(2J+1)= 2.56R \approx 21.3$ J/mol-K, where $R$ refers to the gas constant.  This corresponds to full occupation of all angular momentum multiplets ($2J+1= 13$ for Tb ions)-- which has been reported in other RE magnetic compounds \cite{KekeFeng2022,Raju1999,Takeda2000}.  
 This implies that all 13 eigenstates, with or without degeneracy in the crystal electric field (CEF) levels, are thermally accessible and equally populated for $T\ge70$ K. In other words,  all CEF  levels lie below   $\sim 6$ meV.  Such low energy scales are expected to have little interference with the Fermi surface properties,  but are likely to influence the magnetic energetics among the local Tb moments in the ordered states, potentially contributing  to the complexity of the $T-H$ phase diagram.

\begin{figure*}[t]
\includegraphics[width=\linewidth]{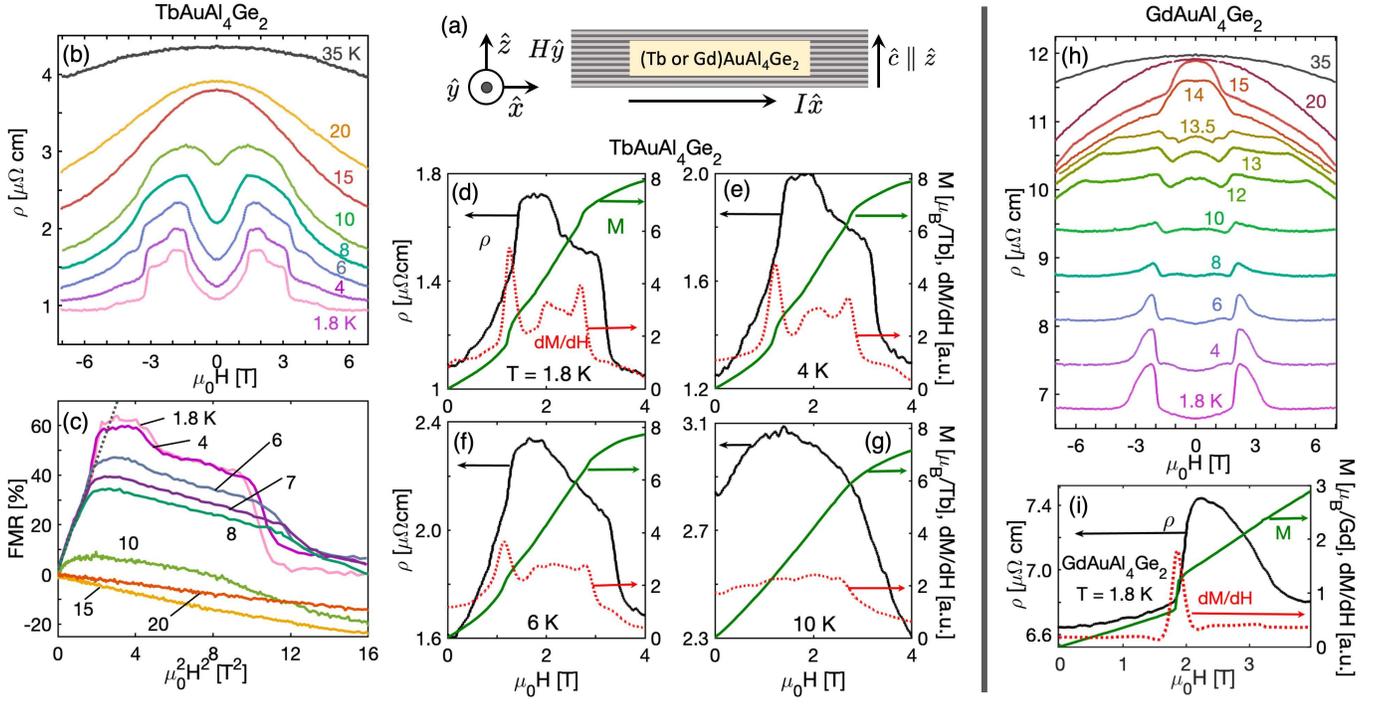}
\caption{ (a)  The schematic diagram of the MR measurement in the $\Hpab$ configuration: current $I$  and field $H$ are both in the $ab$ plane but perpendicular as shown,\iee~$I\hat x \parallel \langle100\rangle$ and  applied field $H\hat y$. 
(b) MR measured at different fixed temperatures from 1.8 K to 35 K.  
(c) The fractional MR plotted against $H^2$, revealing  the initial increase is due to orbital MR where  the charge carrier's motion is  slicing through the Tb-layers. The low field  linear slope (dotted line) shows little $T$ dependence  until the temperature reaches $\ttwo$.  
(d-g) Detailed field dependence  of $\rho(H)$, $\mab$ and its derivative $\dmdh$  is compared at  $T=1.8$, 4, 6 and 10 K respectively.  
(h) The MR $(H)$ of \gaag{} in the same  geometry is plotted at different fixed temperatures. 
(i) Comparison of $\rho(H), \mab(H)$ and $\dmdh$ in \gaag{} at $T=1.8$ K.  Note that  MR decreases with field, even without a saturation of magnetization. 
}
\label{fig:mrab}
\end{figure*}

In order to highlight the effect of the magnetic field on the specific heat, we define $\Delta\cmag (T,H) \equiv  \cmag(H, T) - \cmag( H=0)$  and plot $\Delta\cmag/T$ it in color scale as a function of $H$ and $T$, as shown in  Fig. \ref{fig:hc}(c). 
We note that $\cmag/T = \partial \smag/\partial T$, so a peak in $\Delta\cmag(H)/T$  corresponds to excess magnetic entropy $\smag$ caused by the applied field.

 The dotted lines  show contours of $\Delta\cmag/T$ as it decreases with increasing temperature. These contours are also displayed in our phase diagram in Fig. \ref{fig:magpha}(b).
 For $T<4$ K, we observe a large enhancement of $\Delta\cmag$ that begins at the FM1 region, is centered at the beginning of the FM2 region, and terminates at the beginning of the SP region. 
 This suggests that the ordered state in this region has additional internal degrees of freedom compared to the surrounding areas, which has been observed in other RE metallic compounds \cite{Fushiya2014, Lucas2017}.
Similar behavior was seen for the prototypical Skyrmion lattice material MnSi~\cite{Bauer2013}  as well as in MnGe  as an enhancement of the field dependent thermopower \cite{Fujishiro2018}.  The  enhancement of $\Delta\cmag$ suggests  the formation of  non-trivial spin-textures such as skyrmions or topological bubbles \cite{Shinjo2000, Vistoli2019, Wulferding2017} in this region of the phase diagram.

\subsection {Magnetoresistance }
\subsubsection {$\Hpab$ }
\label{sec:mrab}

Next, we investigate the field dependence of the magnetoresistance (MR) of \taag{} (Tb1142)  in the $\Hpab$ configuration. 
The measurement geometry is illustrated in Fig. \ref{fig:mrab}(a): 
the current  flows along $\hat x$ and the field is applied along $\hat y$.  The  $xy$-plane lies  within the $ab$-plane of Tb1142 with $\hat x \parallel \langle100\rangle$ and  $\hat z$ parallel to the $c$-axis. The voltage contacts are separated along $\hat x$. 

Fig. \ref{fig:mrab}(b) shows the magnetoresistance for Tb1142 at several fixed temperatures for $\Hpab$. Starting from $T=1.8$ K, $\rho(H)$ exhibits a quadratic rise  as a function of $H$ in the  low-field SR region, consistent with conventional orbital MR.  
To obtain further insight we plot the fractional MR (FMR) as  a function of  $H^2$, where the FMR is defined as $\frac{\Delta\rho(H)}{\rho_0}\times 100\%$, with $\rho_0  = \rho(H=0)$ and $ \Delta\rho(H) = \rho(H)-\rho_0 $.  The quadratic dependence in low field persists in $T\lesssim\ttwo$, where the slope marked with a dashed line remains unchanged.  We calculate the effective carrier mobility 
from this slope paired with the carrier density estimated from the high field Hall resistivity to find  $\nu \approx 4.7\times 10^4$ cm$^2$/V$\cdot$s, which is consistent with the measured low resistivity.
In this  particular measurement configuration the carrier cyclotron motion  lies in the $xz$-plane (Fig. \ref{fig:mrab}(a)), and thus,  for a spherical Fermi surface, we estimate the cyclotron radius $r_C = \frac{mv_F}{e\muh}$  to be on the order of a few  hundreds of nm, at $\muh = 2 $ T, which corresponds to  100s of  Tb-layers. Hence, we expect that the interlayer ordering  pattern would make considerable impact on the  MR in addition to  the ordering within individual Tb layers.
 Detailed  investigation on  the properties of magnetic ordering (e.g. neutron scattering ) would shed  more light on this.

Increasing the field further, we observe some detailed structure consisting of a plateau followed by slow and step-like decreases alternating in the FM1 and FM2 regions. Upon entering the SP region, $\rho(H)$ gradually decreases (as $\mab(H)$ increases) as the magnetic moments begin to uniformly align with the applied field. The spins are polarizing but are not fully polarized and the field continues to raise the magnetization and lower the magnetoresistance suppressing spin fluctuations. 

In panels (d-g),  we compare  detailed features in  $\rho(H)$ (left axes) and  $\mab(H)$ and $\dmdh (H)$ (both on right axes) at $T=1.8$,  4, 6, and 10 K, respectively. 
For  $T=1.8$ K [panel  (d)], as the field increases from zero the curvature of $\rho(H)$ becomes stiffer than $H^2$ to reach a plateau, followed by subsequent decreases with two sharp drops. 
The field locations of the plateau and two drops are marked in diamond symbols. in Fig. \ref{fig:magpha}(b). 
These points coincide with features in $\dmdh(H)$ and help to highlight the FM1 and FM2 regions in the phase diagram. 
All features in both $\rho(H)$ and $\mab(H)$ begin to smear with increasing temperature above 6 K and are mostly wiped out by $T=10$ K, just above $\tone$. 
Unlike  the orbital MR induced by the Lorentz force in  the SR region,  the increases and decreases in MR in FM1 and FM2 are likely to  arise from the RKKY interaction of carriers with the magnetic state, reflecting a field-dependent change in the carrier relaxation time, $\tau$. 
The field dependence of $\tau(H)$ is determined by  the interaction  between the Tb moments and conduction electrons via a contact  exchange interaction,  $J_{\rm c-f}$ \cite{LYe2017}.  From the  estimation of $r_C$  at low field,  $\tau(H)$ is expected to depend on the ordering arrangement both within and between Tb layers.

For $T>\ttwo$, Tb1142 enters the wAFM region where  the  suppression of spin scattering by field becomes  more effective as the resistivity at high field is much lower than $\rho_0$ [panel (g)]. This indicates strong spin fluctuations at zero field and hence a weakly ordered state. In this region, the spin interactions responsible for generating the anomalies at low temperatures are destabilized by thermally enhanced spin fluctuations and geometric frustration.

\subsubsection{Comparison with {\rm{\gaag}}}

To gain insight, we compare the in-plane field dependence of the MR   with that of  isostructural \gaag (Gd1142). The Gd analog exhibits   three subsequent magnetic orderings at  $T_{N1,2,3}=17.8, 15.6$, and 13.6 K, identified by features in $\chiab(T)$ (for details, see Ref. \cite{KekeFeng2022}). 
Unlike Tb1142, the magnetization of Gd1142 reveals significantly less anisotropy. This smaller anisotropy in Gd1142 is due to the zero orbital angular moment of the Gd$^{3+}$ ions (with $J=7/2,~L=0,$ and $S=7/2$). At $T=1.8$ K, the $\mab$ of Gd1142 does not get saturated up to $\muh = 7$ T. 

Despite the smaller magnetic anisotropy, Fig. \ref{fig:mrab}(h) shows a featured $\rho(H)$ for $\Hpab$ that evolves as a function of  temperature. 
Below 10K, the field dependence for Gd1142 is simpler yet similar to that of Tb1142. 
The multiple-featured horn shape observed in Tb1142 is replaced with a smooth, featureless horn in Gd1142. In both samples, the evolution with increasing temperature is similar. 
As in Tb1142, the onset of the horn-like region in Gd1142 corresponds to a metamagnetic-like increase in $\mab(H)$ at 1.9T (at 1.8 K). 
We highlight the coincidence of the sharp magnetization increase and the onset of the horn in the MR at 1.8 K in Fig. \ref{fig:mrab}(i). 
Unlike Tb1142, there is only one sharp magnetization increase in Gd1142 which corresponds to the onset of the horn feature. 
Above 10K, the low field MR for Gd1142 develops a concave curvature that narrows in field range with increasing temperature. Above $T=12$ K, an additional shoulder-like feature followed by negative MR emerges, gradually transitioning to negative, parabolic MR at all fields above $\tthree = 17.8$ K. At these elevated temperatures, none of the observed subtle features in Gd1142 are reflected in its magnetization. This may indicate that the MR is more sensitive to subtle rearrangements in the magnetic states that leave the bulk magnetization unaltered.

The MRs for $\Hpab$ in both samples bears a striking similarity. Above the ordering temperature, applied fields reduce scattering and decrease the resistivity. Below their ordering temperatures, both samples show an MR with a horn feature that coincides with a sharp metamagnetic-like transition \cite{KekeFeng2022}. This feature is present at low temperatures and disappears before reaching the highest ordering temperature and  is embedded with subtle changes likely corresponding to intermediate magnetic states.   
It is also interesting to compare the high field behavior at low temperature  in Gd1142 [panel (i)], and Tb1142 [panel (d)] : for Gd1142, MR decreases and flat  even  in absence of  the saturation of magnetization, while the MR drop happens  at higher field than where $\dmdh$ drop occurs.  This implies  that Tb1142 undergoes much stronger spin fluctuations and hence higher degrees of magnetic frustration than Gd1142, which may be related to   the different angular momentum states of two ions. 
Based on our observations in Tb1142 and Gd1142,  we expect the broader family of $Ln$-1142 compounds to show varied, interesting correlated magnetic and electrical behavior.  
The  intricate and diverse field dependence of the MR  and its origin is of great interest in order to understand the RKKY interaction for different ions  with distinct total angular momentum states.

\subsubsection{$\Hpc$ }
\label{sec:mrc}

\begin{figure}[t]
\includegraphics[width=\linewidth]{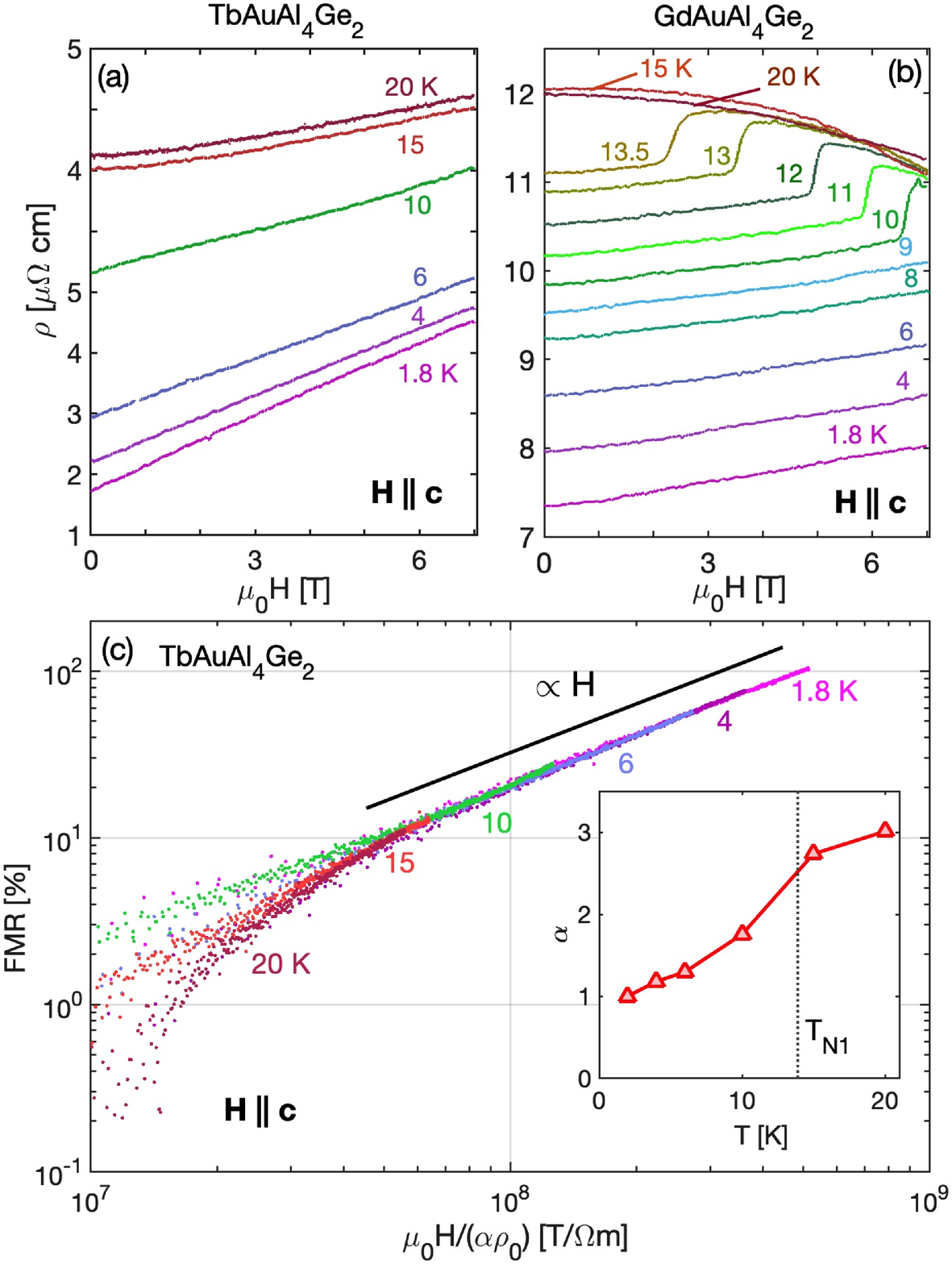}
\caption{(a) Positive MR(H) of \taag{} in the $\Hpc$ configuration, where the current direction remains along $\hat x$ but the field is rotated so that $H \parallel \hat c$ (Fig.\ref{fig:mrab}(a)). 
MR exhibits strikingly linear $H$ dependence, below  $\ttwo$.  
 (b) MR(H) of \gaag~with $\Hpc$. The similar linear positive MR persists  up to 10 K. 
 At 10K, a sharp, step-like increase occurs at high field ($\HKc$) and moves toward smaller fields with increasing temperature. Above $T_{\rm N3} = 13.6 K$, the MR becomes negative and parabolic. 
 (c) Modified Kohler plot of the FMR vs. $\mu_0H/(\alpha\rho_0)$ for \taag, where an additional scaling factor, $\alpha$, is introduced to scale the zero field resistivity and collapse the data. The inset plots this scaling factor vs temperature. As $T$ is lowered,   $\alpha$ beings to  decrease  below $\tone$  and approach unity below $\ttwo$, indicating a change of the scattering mechanism in $\ttwo<T<\tone$. }
\label{fig:mrc}
\end{figure}

In the $\Hpc$ configuration with $I\parallel \hat x$ (measuring the same longitudinal resistivity), we observe starkly contrasting behavior. Fig. \ref{fig:mrc} (a) and (b) display the field dependence of the MR for Tb1142 and Gd1142, respectively. In Tb1142, for all measured temperatures to 20K the MR with $\Hpc$ remains positive. For $T = 1.8-6$ K, the MR is strictly field-linear. Above $\ttwo$, the low field MR starts to develop non-linear curvature, but remains positive. 

In Gd1142, we observe similar $H$-linear MR for $T = 1.8-9$K. Between 10 and 13.5 K the MR remains linear below a step-like transition that appears at high field ($\HKc$) and moves towards lower fields with increasing temperatures. We note that no corresponding feature was observed in $\mc(H)$ at $H_{\rm Kc}$ (data shown in Ref.\cite{KekeFeng2022}). For 15 K and above, the MR turns negative (merging with the behavior above the emergent step feature). 

To disentangle the carrier scattering mechanisms when $\Hpc$ in Tb1142, we  adopt a phenomenological modification to the standard Kohler plot [Fig.~\ref{fig:mrc} (c)]. We introduce a field-independent but  $T$-dependent scaling factor, $\alpha(T)$, which multiplies $\rho_0 (T)$ to  cause all MR curves  to collapse. 
The scaling factor remains near 1 for $T<\ttwo$ where  linear MR is most prominent, then rapidly increases towards a value of 3 as $T$ increases through $\tone$. This is interpreted as  a 3-fold increase in scattering with field  in the higher $T$ wAFM region, compared to below the ordering temperature.

While the mechanism responsible for the linear MR in Gd1142 and Tb1142 is not immediately obvious, it is clear that the linear MR is not directly correlated to the spin degree of freedom. Gd1142 and Tb1142 host  different degree of  anisotropies with correspondingly different magnetizations for $\Hpc$ (For Tb1142, $\mab$ is more than  five times larger than $\mc$, while for Gd1142 it is only 1.4 times larger), yet both samples show similarly sized linear MR. Linear MR has been shown to originate from electronic origins as a result of certain Fermi surface ~\cite{ Oomi1997, Budko1998, Lucas2017} or band structure properties ~\cite{LYe2018}. 
Some $f$-electron metallic systems are found to  exhibit linear MR over large field ranges  that has been connected to modification of high curvature Fermi surfaces due to zone-folding energy gaps ~\cite{Tsuda2018, Feng2019,  Kolincio2020}, resulting from the formation of charge or spin density waves. 
For Tb1142 and Gd1142, linear MR is only seen for $T<\ttwo$, which may suggest that incommensurate AFM order may introduce gapped Fermi surface regions leading to  linear MR for $\Hpc$. Meanwhile the $ab$-plane behavior is dominated by field-induced transitions and the corresponding changes in spin-scattering. 
Charge density wave formation has  been observed in other lanthanide compounds with unambiguous signatures in the diffraction pattern as well as a shoulder-like increase of the resistivity at the Peierls transition temperature \cite{Stavinoha2018, Kolincio2020}.  This motivates further studies to understand this behavior and to determine whether it is present in the chemical analogs.

\section {Summary}

We have reported  the heat capacity and field-direction dependent magnetoresistance  results in the metallic rare-earth layered triangular antiferromagenetic compound \taag. 
A small range of applied fields along the $ab$-plane drives the system through multiple magnetic states as reflected in the magnetoresistance and magnetization, resulting in a rich magnetic phase diagram where the energy scales of these states are narrowly separated.
 Simultaneously, a region of enhanced magnetic specific heat suggests the possibility of non-trivial spin textures hosted in the low-$T$ regions of FM1 and FM2. 
 For $\Hpc$,  positive linear MR emerges over a wide field range below the ordering temperature, which  is also observed in the isostructural, more magnetically-isotropic \gaag. 
 We suspect that the linear magnetoresistance is electronic in origin:  one possibility is that the Fermi surface in both compounds undergoes reconstruction due to density wave physics or other electronic correlation effects. 
 Taken together, these measurements motivate further work to clarify the details of the magnetic states in TbAuAl$_4$Ge$_2$, such as neutron scattering, magnetic force microscopy, or Lorentz tunneling electron microscopy.  It is also of interest to examine the variation of  our observations,  by varying the lanthanide and transition metal ions in the chemical analogs.

\begin{acknowledgments}

M.L. and I.A.L  were supported by  the U.S. Department of Energy, 
Basic Energy Sciences, Materials Sciences and Engineering Division under Award No.~DE-SC0021377. R.B. and K.F were supported by the National Science Foundation through NSF DMR-1904361. R.D. acknowledges the support from the  NSF  Research Experiences for Undergraduates program of 2021. 
A part of electrical transport measurements were performed at the National High Magnetic Field 
Laboratory, which is supported by National Science Foundation Cooperative Agreement 
DMR- 1644779 and the State of Florida.

\end{acknowledgments}


%

\end{document}